\documentclass[aps,prl,twocolumn,superscriptaddress,showpacs, amsmath, amssymb]{revtex4}
\usepackage[dvips]{graphicx}
\usepackage{dcolumn}
\usepackage{bm}
\usepackage{color}

\begin{document}

\preprint{APS/123-QED}

\title{Inelastic Collisions in Optically Trapped Ultracold Metastable Ytterbium}

\author{A. Yamaguchi}
\affiliation{Department of Physics, Graduate School of Science, Kyoto University, Kyoto 606-8502, Japan}
\author{S. Uetake}
\affiliation{CREST, Japan Science and Technology Agency, Kawaguchi, Saitama 332-0012, Japan}
\author{D. Hashimoto}
\affiliation{Department of Physics, Graduate School of Science, Kyoto University, Kyoto 606-8502, Japan}
\author{J. M. Doyle}
\affiliation{Harvard-MIT Center for Ultracold Atoms, Cambridge, Massachusetts 02138, USA}
\affiliation{Department of Physics, Harvard University, Cambridge, Massachusetts 02138, USA}
\author{Y. Takahashi}
\affiliation{Department of Physics, Graduate School of Science, Kyoto University, Kyoto 606-8502, Japan}
\affiliation{CREST, Japan Science and Technology Agency, Kawaguchi, Saitama 332-0012, Japan}
\date{\today}

\begin{abstract}
We report measurement of inelastic loss in dense and cold metastable ytterbium (Yb[$^3P_2$]). Use of an optical far-off-resonance trap enables us to trap atoms in all magnetic sublevels, removing multichannel collisional trap loss from the system. Trapped samples of Yb[$^3P_2$] are produced at a density of 2$\times$10$^{13}$ cm$^{-3}$ and temperature of 2 $\mu$K. We observe rapid two-body trap loss of Yb[$^3P_2$] and measure the inelastic collision rate constant 1.0(3)$\times$10$^{-11}$ cm$^3$s$^{-1}$. The existence of the fine-structure changing collisions between atoms in the $^3P_2$ state is strongly suggested.
\end{abstract}

                            
\pagestyle{empty}
\maketitle

\newpage 

There is increasing interest in ultracold two-electron atoms \cite{Clock works, PA works}, such as the alkaline earth metals (e.g. Ca and Sr) and Yb. In particular, novel characteristics of the metastable $^3P_2$ atoms have recently attracted attention, both for applications and for the study of their collisional properties \cite{Derevianko Julienne}. These atoms are set apart from the more commonly studied alkali metal atoms because collisions between $^3P_2$ atoms are intrinsically anisotropic. Recent theory has investigated the effects of this anisotropy, including its interplay with magnetic field effects, which enable novel control of the scattering length \cite{Santra PRA}, and multichannel collisions due to a strong coupling among the partial waves of relative motion \cite{Kokoouline PRL, Neon works}. Also, the magnetic dipole-dipole interaction between $^3P_2$ atoms is 9 times larger than that between alkali metal atoms. This has led to theoretical predictions such as novel quantum phases and use in quantum information systems \cite{Lewenstein, Derevianko Quantum Computing}. 

In order to move toward study of these new possible features of $^3P_2$ atoms, several laboratories have realized laser cooling and trapping of metastable two-electron atoms. Ca and Sr atoms decaying to the $^3P_2$ state from the $^1P_1$ state, which is the upper state in the $^1S_0$$\leftrightarrow$$^1P_1$ magneto-optical trap (MOT) transition, have been successfully trapped in a magnetic trap \cite{3P2 MT}. Also, a MOT operating on the $^3P_2$$\leftrightarrow$$^3D_3$ transition has been used to load a magnetic trap \cite{3P2 MOT Hemmerich}. In spite of successes of these approaches, evaporative cooling of $^3P_2$ atoms in a magnetic trap to reach Bose-Einstein condensation (BEC) turned out to be unsuccessful due to trap loss caused by strong multichannel collision processes \cite{Kokoouline PRL}. More recently, a similar large inelastic collision rate in Ca[$^3P_2$, $m_J$=2] was observed \cite{3P2 Multichannel Hansen}.  

The loss induced by multichannel collisions in a magnetic trap can be overcome by employing, instead of a magnetic trap, an optical far-off-resonance trap (FORT). According to Ref. \cite{1S0-3P2}, the FORT wavelength can be chosen so that atoms in every magnetic sublevel of the $^3P_2$ state can be trapped with the same strength. As a result, although multichannel collisions can still occur, which distributes the atoms over the different magnetic sublevels, they will not lead to trap loss. Thus, any trap loss observed in such a trap must be due to a different physical mechanism. Study of these collisional properties is crucial to understanding the physics of these important class of atoms and states.

In this paper, we report both the experimental realization of optical trapping (FORT) of ultracold $^{174}$Yb[$^3P_2$] atoms at high atom number density and the quantitative measurement of inelastic collisions. In contrast to previous methods, we first trap atoms and perform evaporative cooling using Yb[$^1S_0$] in the FORT. We optically excite Yb[$^1S_0$] to the $^3P_2$ state to obtain ultracold trapped Yb[$^3P_2$], achieving a number density of 2$\times$10$^{13}$ cm$^{-3}$ at a temperature of 2 $\mu$K with phase space density (PSD) of 0.01. Our newly achieved atom number density is larger than the previous work by an order of three \cite{3P2 MOT Hemmerich}. We also measured a large two-body inelastic collision rate in the FORT which we interpret as fine-structure changing collisions in this ultracold temperature regime. Although fine-structure changing inelastic collisional properties have previously been investigated for Mg[$^3P_j$], O[$^3P_j$], Sc[$^2D_j$], and Ti[$^3F_j$] colliding with closed shell atoms \cite{Fine structure changing collision Mg,Fine structure changing collision O,Fine structure changing collision Ti,Fine structure changing collision Ti Sc theory}, they had not been seen in collisions between $^3P_2$ atoms. While the recent experiment of magnetically trapped Ca atoms studied multichannel collisions between $^3P_2$ atoms and discussed the possibility of the fine-structure changing process \cite{3P2 Multichannel Hansen}, we believe that our work is the first definite experimental measurement of this process between $^3P_2$ atoms.

\begin{figure}
\includegraphics[width=\linewidth]{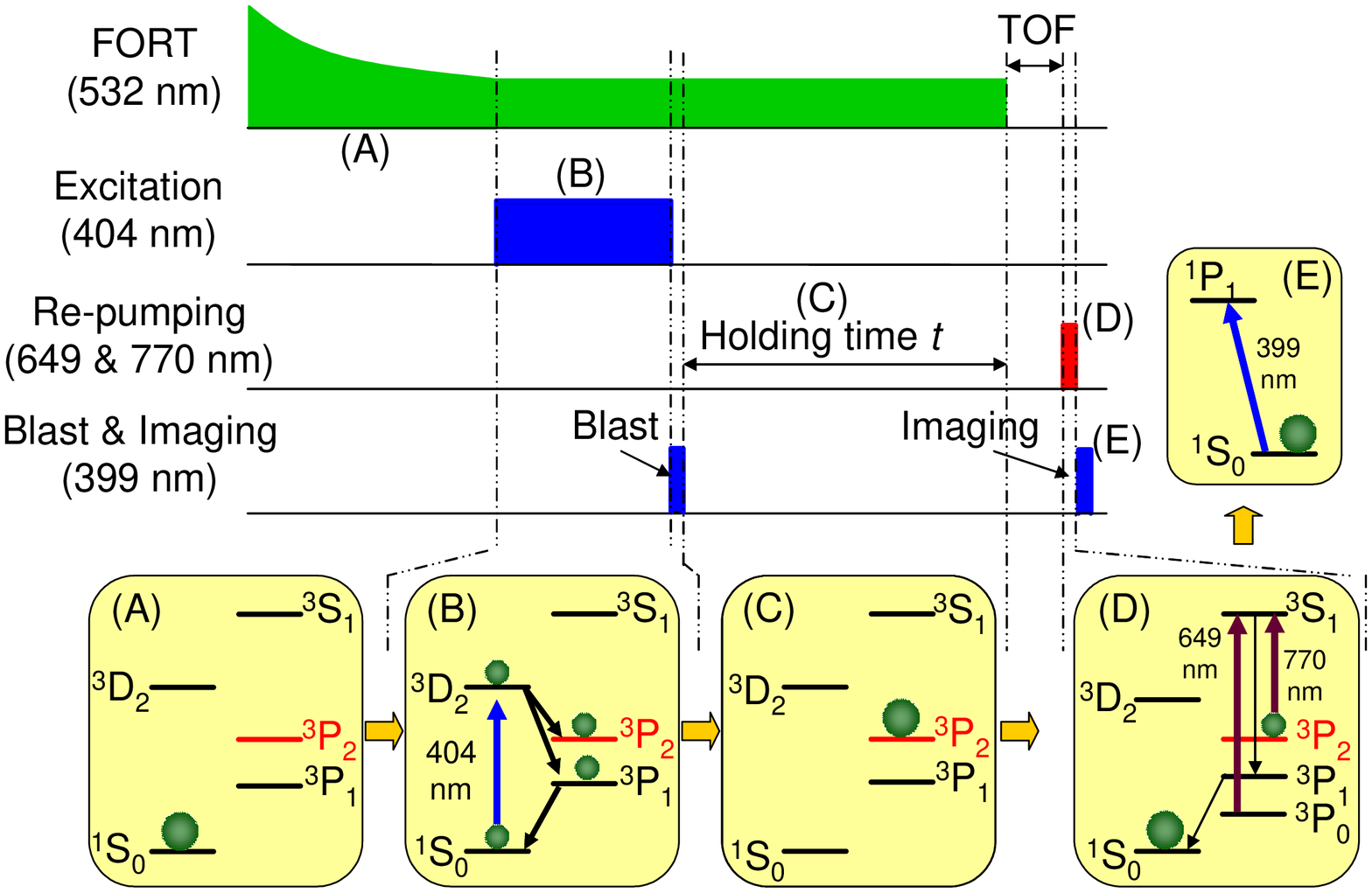}
\\
\quad\\
\includegraphics[width=6cm]{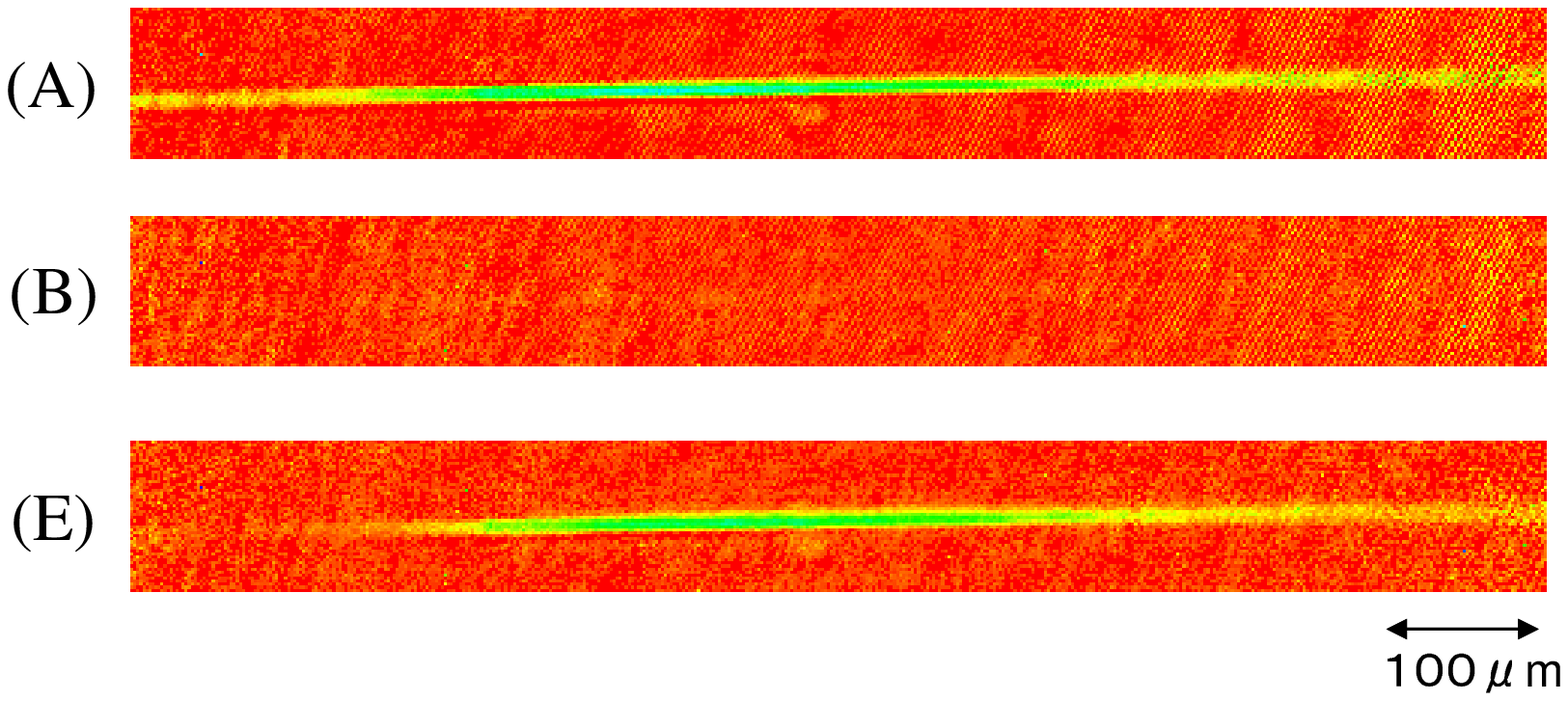}
\caption{\label{fig:TimeSequence and CCD images}(color online). Experimental procedures to excite atoms to the $^3P_2$ state. Absorption images of atoms in the ground state at stages A, B, and E are also shown below. (A): Yb atoms trapped in a MOT are transferred to the FORT and cooled by evaporative cooling. (B): Then atoms are optically excited to the $^3P_2$ state in the FORT via the intermediate $^3D_2$ state and subsequent spontaneous decay $^3D_2$$\to$$^3P_2$. Atoms remaining in the ground state are blew by a short blast laser pulse just after the excitation. Thus no atoms in the ground state are observed in the CCD image at this stage. (C): Metastable atoms are left in the FORT during the holding time $t$. (D): After a TOF time, metastable atoms are repumped to the ground state by 770 nm ($^3P_2 \leftrightarrow ^3S_1$) and 649 nm ($^3P_0 \leftrightarrow ^3S_1$) laser pulses of 100 $\mu$s and (E) detected by the absorption imaging.}
\end{figure}

Our typical procedure to prepare and detect atoms in the $^3P_2$ state is summarized in Fig.\ref{fig:TimeSequence and CCD images}. At the first stage A, we prepare cold $^{174}$Yb atoms in the ground state in the FORT by almost the same method as that of our previous works \cite{Yb BEC, Yb FD} which is briefly summarized here. With about 10 sec loading, we typically collect 10$^7$ Yb atoms in the ground state at a temperature below 50 $\mu$K by the MOT using the narrow intercombination transition $^1S_0$$\leftrightarrow$$^3P_1$. These atoms are then transferred to a single or crossed FORT created by focused diode-pumped solid state 10 W-lasers at 532 nm. After carrying out evaporative cooling by gradually reducing the trap depth, we typically have 10$^6$ atoms at a temperature of 30 $\mu$K in the single FORT. When we use a crossed FORT configuration and perform evaporative cooling, a temperature of atoms reaches below 1 $\mu$K.

\begin{figure}
\includegraphics[width=8cm]{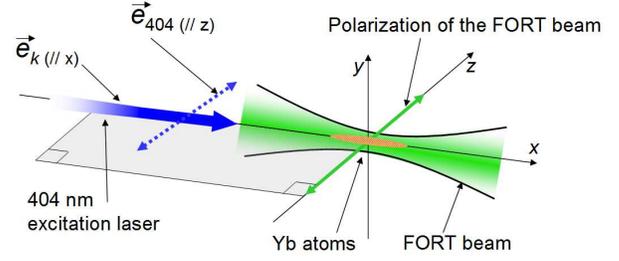}
\caption{\label{fig:3D2 excitation and lightshift}(color online). Experimental setups are schematically shown. The $\vec{\textbf{e}}_{404}$ and $\vec{\textbf{e}}_{k}=\vec{\textbf{k}}/|\vec{\textbf{k}}|$ are the polarization and the wavenumber vector of the 404 nm excitation laser.}
\end{figure}

At the next stage B, we optically excite atoms to the $^3P_2$ state in the FORT. We use the intermediate $^3D_2$ state for the excitation (transition linewidth: 350 kHz \cite{Berkeley 1S0-3D2 404nm}) and the subsequent spontaneous decay to the $^3P_2$ state with the lifetime 460 ns of the $^3D_2$ state. Although some atoms in the $^3D_2$ state can also decay to the $^3P_1$ state, they immediately decay to the ground state with the lifetime 875 ns of the $^3P_1$ state and are re-excited to the $^3D_2$ state. Thus, after iterating this excitation cycle for several times, we can transfer all atoms to the $^3P_2$ state typically within 5 ms. In order to efficiently excite atoms, we stabilize a blue GaN laser diode (404 nm) by an external cavity laser diode system combined with an optical feedback technique \cite{OpticalFeedback}. The resulting linewidth is narrowed to 1 MHz for 0.5 s. The excitation laser is superposed with the FORT laser (see Fig. \ref{fig:3D2 excitation and lightshift}) and the peak intensity at atoms reaches 1.5 W/cm$^2$. Just after the excitation, we irradiate a short blast pulse which blows remaining atoms in the ground state by using the strong $^1S_0$$\leftrightarrow$$^1P_1$ transition.

The $^1S_0$$\leftrightarrow$$^3D_2$ transition (404 nm) is the electric quadrupole (E2) transition. In this experiment, as shown in Fig. \ref{fig:3D2 excitation and lightshift}, we set $\vec{\textbf{e}}_{404} \parallel z$ and $\vec{\textbf{e}}_{k} \parallel x$, where $\vec{\textbf{e}}_{404}$ and $\vec{\textbf{e}}_{k}=\vec{\textbf{k}}/|\vec{\textbf{k}}|$ are the polarization and the wavenumber vectors of the 404 nm excitation laser. Thus atoms are excited to the $^3D_2$, $m=\pm 1$ states due to the selection rule and subsequently decay to all the magnetic sublevels of the $^3P_2$ state with the ratio $3:1:2$ for $|m|$=0, 1, and 2 sublevels of the $^3P_2$ state, respectively. 

At the stage C, we hold metastable atoms in the FORT for a time $t$. Then, at the stage D, in order to measure the temperature and the number of atoms in the $^3P_2$ state with a time-of-flight (TOF) technique, we rapidly repump all the metastable $^3P_2$ atoms to the ground state after a TOF time. A 770 nm ($^3P_2$$\leftrightarrow$$^3S_1$)  resonant pulse excites atoms from the $^3P_2$ state to the $^3S_1$ state, from which all the $^3P_2$, $^3P_1$, and $^3P_0$ states are populated through spontaneous decay. By simultaneous application of a 649 nm ($^3P_0$$\leftrightarrow$$^3S_1$) resonant pulse, all atoms return through the $^3P_1$ state to the ground state where we can use an absorption imaging technique using the strong cyclic $^1S_0$$\leftrightarrow$$^1P_1$ transition at the stage E. Since 100 $\mu$s duration for the repumping procedure is short enough compared with a typical TOF time, we can safely regard that the observed atomic distribution precisely reflects that of atoms in the $^3P_2$ state.

\begin{figure}
\includegraphics[width=7.5cm]{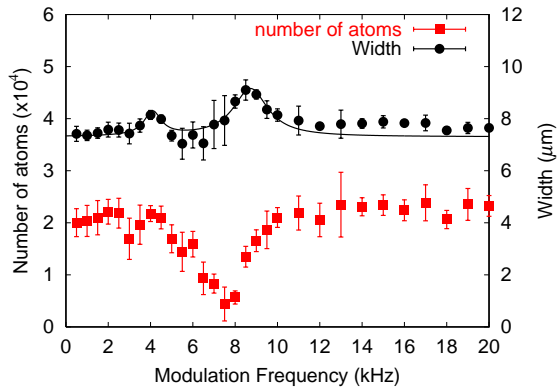}
\caption{\label{fig:parame-yoko} (color online). Measurement of the trap frequency of the $^3P_2$ state by using the parametric resonance technique. Number of trapped atoms and the width of the atomic cloud are plotted as a function of the modulation frequency. }
\end{figure}

We first measured a trap frequency for the atoms in the $^3P_2$ state in the single FORT by using a parametric resonance technique \cite{parametric resonance}, where atoms are heated at the modulation frequency $\omega = 2 \omega_{\mathrm{trap}}/n$ with $\omega_{\mathrm{trap}}$ the trap frequency and $n$ an integer. To this end, the FORT power was modulated at the stage C (see Fig. \ref{fig:TimeSequence and CCD images}). Here we took special care about the cancellation of a magnetic field since the light shifts due to the FORT beam in the presence of a large magnetic field are quite sensitive to the relative orientation between the magnetic field and the FORT polarization \cite{Ido 3P1}. We measured the number of remaining atoms and the width of the atomic cloud, which are shown in Fig. \ref{fig:parame-yoko}. Two resonance signals corresponding to $n$=1 and 2 are clearly observed at 8.6 and 4.3 kHz, respectively, while the trap frequency of the $^1S_0$ state in this condition is 3.9 kHz. The resonance frequency is determined by fitting a Lorentzian function to the width of atomic clouds. Since all the magnetic sublevels of the $^3P_2$ state are populated in this measurement and in fact we experimentally confirmed  that the atoms in every sublevel are trapped, their trap frequencies can be thought to coincide with each other within the resolution of this measurement. This result is consistent with the measurement of polarizabilities of magnetic sublevels of the $^3P_2$ state using the ultranarrow $^1S_0$$\leftrightarrow$$^3P_2$ transition in our different work, which indicates that the difference of the trap frequency between magnetic sublevels is less than 1 kHz \cite{1S0-3P2}. Thus, in the following discussion, we consider that the trap depth of all magnetic sublevels of the $^3P_2$ state is the same. With this information on the trap, we can accurately estimate a density.

We successfully obtain high density ultracold metastable Yb[$^3P_2$] by exciting pre-cooled Yb[$^1S_0$] in a crossed FORT. These atoms are prepared at a density
of 2$\times$10$^{14}$ cm$^{-1}$ and at a temperature of less than 0.7 $\mu$K. During excitation from the ground state to the $^3P_2$ state (through the intermediate $^3D_2$ state), the atoms suffer from heating due to the spontaneous decay. As a result, the density of atoms decreases and the temperature increases, resulting in density $n = 2\times10^{13}$ cm$^{-3}$, temperature $T$ = 2.0 $\mu$K and PSD = 0.01. This is the highest density and lowest temperature ever achieved for $^3P_2$ atoms. Although in principle PSD could be increased by direct evaporative cooling of the Yb[$^3P_2$] atoms, we found this impossible due to an inelastic collisional process which we identify as fine-structure changing collisions.

\begin{figure}
\includegraphics[width=7.5cm]{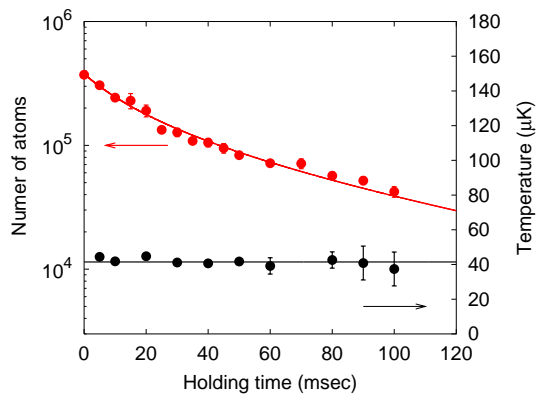}
\caption{\label{fig:decay and temperature}(color online). Number of atoms and temperature of atoms in the FORT are plotted as a function of holding time. The non-exponential rapid decay can be well described by the Eq.(\ref{decay equation}), which implies the large two-body loss rate of metastable atoms. Almost constant temperature during the measured holding time indicates that the system is in the thermal equilibrium. }
\end{figure}

We studied inelastic loss by measuring the number density decay of Yb[$^3P_2$] atoms from a single beam FORT. Figure 4 shows the number of $^3P_2$ atoms in the FORT as a function of time after loading. Note that the (presumably state independent) one-body trap loss lifetime due to background gas collisions is measured to be 15 s, much longer than the observed trap lifetime. A significant feature of the data is the non-exponential decay of atom number along with the constant temperature of $T$ = 41 $\mu$K. Our model for atom loss includes a combination of one-body loss and two-body loss, as embodied in the following equation for atom number versus time.
\begin{equation}
\label{decay equation}
\frac{\mathrm{d}N}{\mathrm{d}t} = -\Gamma N -\beta' N^2,
\end{equation}
where $\Gamma$ is the one-body loss rate and $\beta'$ is the measured two-body atom loss rate. The solid line in Fig. \ref{fig:decay and temperature} for data of the number of atoms in the FORT is a fit by Eq. (\ref{decay equation}). $\beta'$ is related to the density related (volume independent) two-body loss rate coefficient $\beta$ by $\beta = \beta' V_{\mathrm{eff}}$, where $V_{\mathrm{eff}}$ is the effective volume of the atoms. In the present experiment, the volume can be well approximated by a cylinder with the size of the FORT beam \cite{Wieman Loading to optical dipole trap}
\begin{equation}
\label{volume}
V_{\mathrm{eff}} = \pi \omega_0^2 z_{\mathrm{R}} \ln \left( \frac{\eta}{\eta-1} \right) \sqrt{\frac{1}{\eta-1}},
\end{equation}
where $\omega_0$ and $z_{\mathrm{R}}$ are the beam waist and the Rayleigh length of the FORT beam, respectively. $\eta$ $(= U_0/(k_{\mathrm{B}}T))$ is the ratio of the trap depth $U_0$ to the temperature $T$, and $k_{\mathrm{B}}$ is the Boltzmann constant. From the measured trap frequency, the trap depth is found to be $U_0/k_{\mathrm{B}} = $193 $\mu$K for all magnetic sublevels and thus $\eta$ = 4.7 which remains constant throughout the 120 ms holding time.

The two-body collisional loss of atoms from the trap can be due to elastic and inelastic processes, identified as elastic ($\beta_{\mathrm{el}}$) and inelastic ($\beta_{\mathrm{in}}$) collision rates, respectively. Under thermal equilibrium conditions, the relation between the observed two-body decay rate $\beta$ and the elastic $\beta_{\mathrm{el}}$ and inelastic $\beta_{\mathrm{in}}$ collision rate are described by
\begin{equation}
\label{2-body-decay}
\beta_{\mathrm{in}} = \displaystyle\frac{1}{f\gamma + 1}\beta,\quad \beta_{\mathrm{el}} = \displaystyle\frac{\gamma}{f\gamma + 1}\beta, 
\end{equation}
where $f$ is the evaporation fraction and $\gamma \ \equiv \sigma_{\mathrm{el}}/\sigma_{\mathrm{in}}$ is the ratio of the elastic to inelastic cross sections \cite{Doyle low-eta evaporative cooling}. All of these parameters $f$ and $\gamma$ are functions of only $\eta$. Thus, by slightly modifying the method in \cite{Doyle low-eta evaporative cooling}, we can estimate both the inelastic scattering rate $\beta_{\mathrm{in}}$= 1.0(3)$\times$10$^{-11}$ cm$^3$s$^{-1}$ and the elastic one $\beta_{\mathrm{el}}$= 2.3(6)$\times$10$^{-11}$ cm$^3$s$^{-1}$ for the temperature 41 $\mu$K. The elastic binary collision rate can be independently investigated by the cross-dimensional relaxation measurement \cite{Cross Dimensional Collision Relaxation by Wieman} and the result was consistent with the above estimation.

Considering the suppression of multichannel collisional loss in our FORT which can trap atoms in every magnetic sublevel of the $^3P_2$ state with zero magnetic field, the observed inelastic collision rate is anomalously large. Thus a different inelastic collision process -- fine-structure changing collisions -- is strongly suggested. The previous theoretical works revealed details of fine-structure changing transitions in collisions of Mg[$^3P_j$], O[$^3P_j$], Sc[$^2D_j$], and Ti[$^3F_j$] atoms with closed-shell atoms at a high temperature \cite{Fine structure changing collision Mg,Fine structure changing collision O,Fine structure changing collision Ti,Fine structure changing collision Ti Sc theory}. However, there has not been no theoretical work on the fine-structure changing collisions between atoms in the $^3P_2$ state at ultralow temperature achieved in the present work. As our experimental results suggest, further theoretical investigation is warranted. 

In conclusion we have optically trapped high density metastable ytterbium atoms in the FORT. The achieved number density is 2$\times$10$^{13}$ cm$^{-3}$ at a temperature 2 $\mu$K. The non-exponential rapid loss of atoms in the FORT at a constant temperature is observed, and then a large inelastic binary collision rate constant is measured. We interpret this as fine structure changing collisions between $^3P_2$ atoms.

We acknowledge all members of the Yb team at Kyoto, M. Mizoguchi, S. Kato and T. H. Yoon for experimental assistances and insightful discussions. This research was partially supported by Grant-in-Aid for Scientific Research of JSPS (18043013, 18204035), SCOPE-S, and 21st Century COE hCenter for Diversity and Universality in Physicsh from MEXT of Japan. A.Y. acknowledges support from JSPS.

\end{document}